\documentclass[preprint]{revtex4}

\usepackage{color}
\usepackage{graphicx}
\usepackage{amsmath}
\usepackage{amssymb}

\newcommand{\CiteHotAtom}{\cite{SFWM.dLambda.HotAtom1, SFWM.dLambda.HotAtom2, SFWM.dLambda.HotAtom3, SFWM.dLambda.HotAtom4, SFWM.dLambda.HotAtom5, SFWM.dLambda.HotAtom6}}

\newcommand{\CiteSPDC}{\cite{SPDC.Cavity1, SPDC.Cavity2, SPDC.Cavity3, SPDC.Cavity4, SPDC.Cavity5, SPDC.Cavity6, SPDC.Cavity7, SPDC.Cavity8, SPDC.Cavity9, SPDC.Cavity10, SPDC.Cavity11, SPDC.Cavity12, SPDC.Cavity13, LongBiphotonSPDC.Cavity}}

\newcommand{\CiteColdAtom}{\cite{FirstSFWM, SFWM.dLambda.ColdAtoms2, SFWM.dLambda.ColdAtoms3, SFWM.dLambda.ColdAtoms4, SFWM.dLambda.ColdAtoms5, SFWM.dLambda.ColdAtoms6, SFWM.dLambda.ColdAtoms7, SFWM.dLambda.ColdAtoms8, SFWM.dLambda.ColdAtoms9, SFWM.dLambda.ColdAtoms10, SFWM.dLambda.ColdAtoms11, SFWM.dLambda.ColdAtoms12, SFWM.dLambda.ColdAtoms13}} 

\newcommand{\CiteLadderType}{\cite{SFWM.ladder.ColdAtoms1, SFWM.ladder.ColdAtoms2, SFWM.ladder.HotAtom1, SFWM.ladder.HotAtom2, SFWM.ladder.HotAtom3, SFWM.ladder.HotAtom4, SFWM.ladder.HotAtom5}}

\begin{document}
\title{
Generation of sub-MHz and spectrally-bright biphotons from hot atomic vapors with a phase mismatch-free scheme}

\author{Chia-Yu Hsu$^1$} 
\author{Yu-Sheng Wang$^1$} 
\author{Jia-Mou Chen$^1$} 
\author{Fu-Chen Huang$^1$} 
\author{Yi-Ting Ke$^1$} 
\author{Emily Kay Huang$^1$}
\author{Weilun Hung$^1$} 
\author{Kai-Lin Chao$^1$} 
\author{Shih-Si Hsiao$^1$} 
\author{Yi-Hsin Chen$^{2,5}$} 
\author{Chih-Sung Chuu$^{1,5}$}
\author{Ying-Cheng Chen$^{3,5}$}
\author{Yong-Fan Chen$^{4,5}$}
\author{Ite A. Yu$^{1,5,}$}\email{yu@phys.nthu.edu.tw}

\affiliation{
$^1$Department of Physics, National Tsing Hua University, Hsinchu 30013, Taiwan \\
$^2$Department of Physics, National Sun Yat-Sen University, Kaohsiung 80424, Taiwan \\
$^3$Institute of Atomic and Molecular Sciences, Academia Sinica, Taipei 10617, Taiwan\\
$^4$Department of Physics, National Cheng Kung University, Tainan 70101, Taiwan \\
$^5$Center for Quantum Technology, Hsinchu 30013, Taiwan
}

\begin{abstract}
We utilized the all-copropagating scheme, which maintains the phase-match condition, in the spontaneous four-wave mixing (SFWM) process to generate biphotons from a hot atomic vapor. The linewidth and spectral brightness of our biphotons surpass those of the biphotons produced with the hot-atom SFWM in the previous works. Moreover, the generation rate of the sub-MHz biphoton source in this work can also compete with those of the sub-MHz biphoton sources of the cold-atom SFWM or cavity-assisted spontaneous parametric down conversion. Here, the biphoton linewidth is tunable for an order of magnitude. As we tuned the linewidth to 610 kHz, the generation rate per linewidth is 1,500 pairs/(s$\cdot$MHz) and the maximum two-photon correlation function, $g_{s,as}^{(2)}$, of the biphotons is 42. This $g_{s,as}^{(2)}$ violates the Cauchy-Schwartz inequality for classical light by 440 folds, and demonstrates that the biphotons have a high purity. By increasing the pump power by 16 folds, we further enhanced the generation rate per linewidth to 2.3$\times$10$^4$ pairs/(s$\cdot$MHz), while the maximum $g_{s,as}^{(2)}$ became 6.7. In addition, we are able to tune the linewidth down to 290$\pm$20 kHz. This is the narrowest linewidth to date among all single-mode biphoton sources of room-temperature and hot media.
\end{abstract}

\maketitle
\newcommand{\FigOne}{
	\begin{figure}[t]
	\center{\includegraphics[width=85mm]{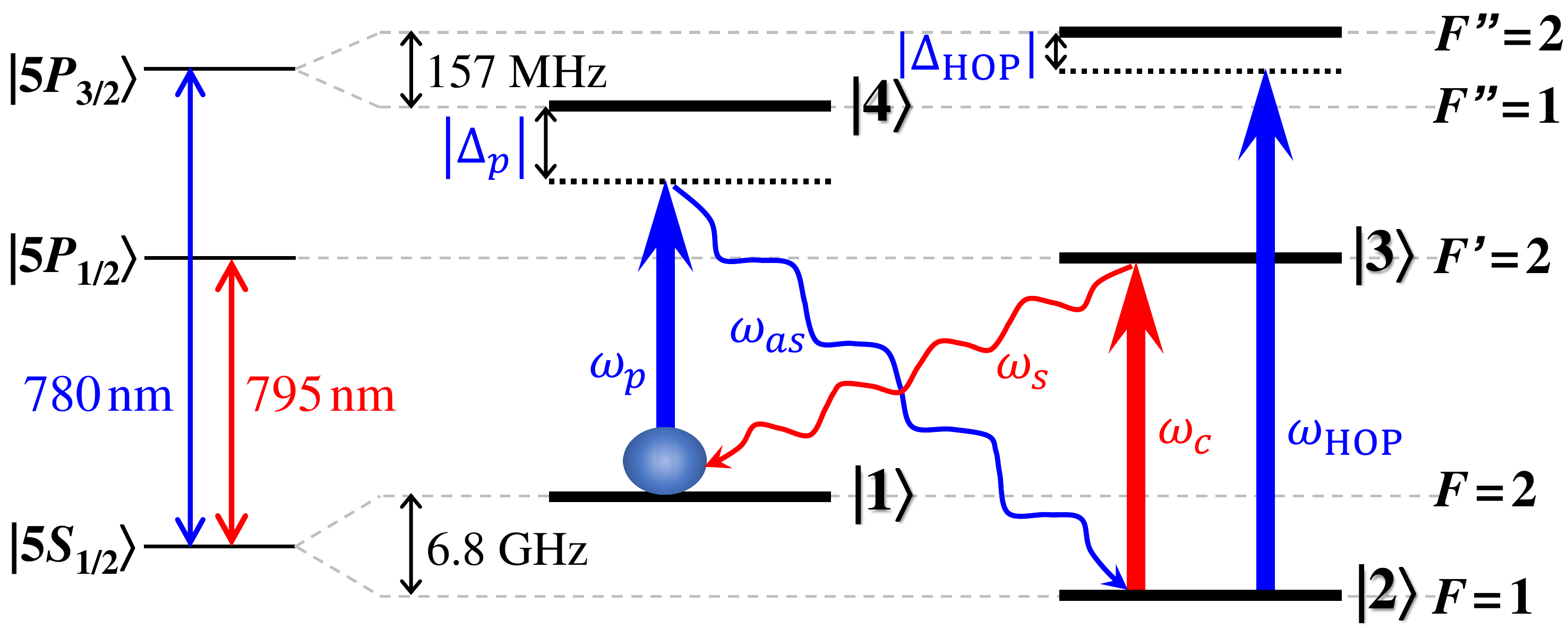}}
	\caption{Relevant energy levels of $^{87}$Rb atoms and the transition diagram. In the SFWM scheme, $\omega_p$, $\omega_{as}$, $\omega_c$, and $\omega_s$ represent the frequencies of the pump field, anti-Stokes photon, coupling field, and Stokes photon, forming the four-photon resonance condition. The detuning of the pump field, $\Delta_p$, is about $-2.0$ GHz. The frequency of the hyperfine optical pumping (HOP) field is denoted by $\omega_{\rm HOP}$, and has the detuning, $\Delta_{\rm HOP}$, of $-80$ MHz.}
	\label{fig:transition_diagram}
	\end{figure}
}
\newcommand{\FigTwo}{
	\begin{figure}[t]
	\center{\includegraphics[width=85mm]{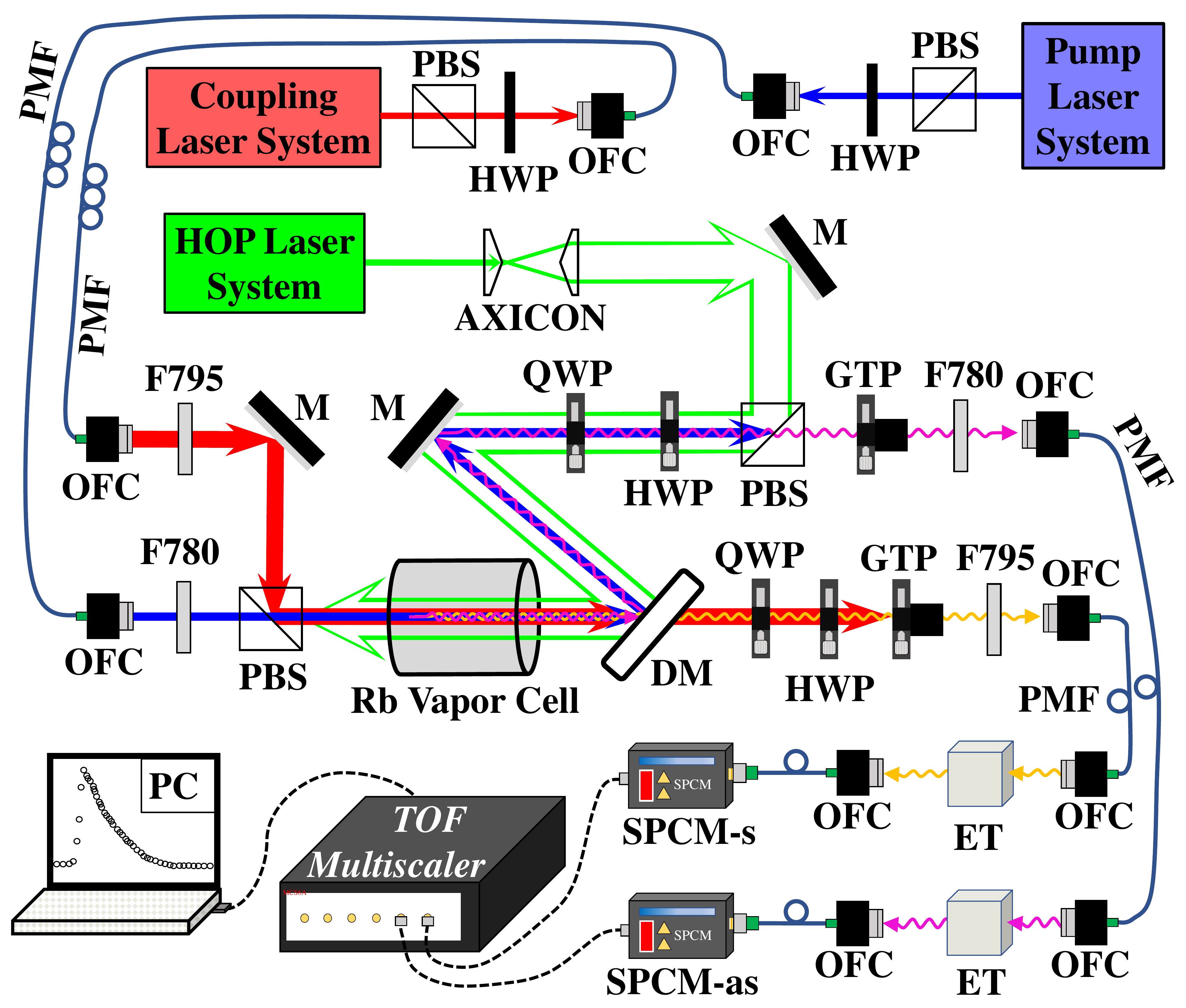}}
	\caption{Experimental setup. Blue, red, and green lines represent the laser beams of the pump, coupling, and HOP fields. Violet and orange wiggly lines indicate the optical paths of the anti-Stokes and Stokes photons. PBS: polarizing beam splitter, HWP: half-wave plate, OFC: optical fiber coupler, PMF: polarization-maintained optical fiber, F780 or F795: a set of three 780 or 795 nm bandpass filters (Thorlabs FBH-800-10 or FBH-780-10) in series, M: mirror, QWP: quarter-wave plate, GTP: Glan-Thompson polarizer, DM: dichroic mirror, ET: a set of two etalons in series, and SPCM-as (SPCM-s): single-photon counting module for anti-Stokes (Stokes) photons.}
	\label{fig:experimental_setup}
	\end{figure}
}
\newcommand{\FigThree}{
	\begin{figure}[t]
	\center{\includegraphics[width=\columnwidth]{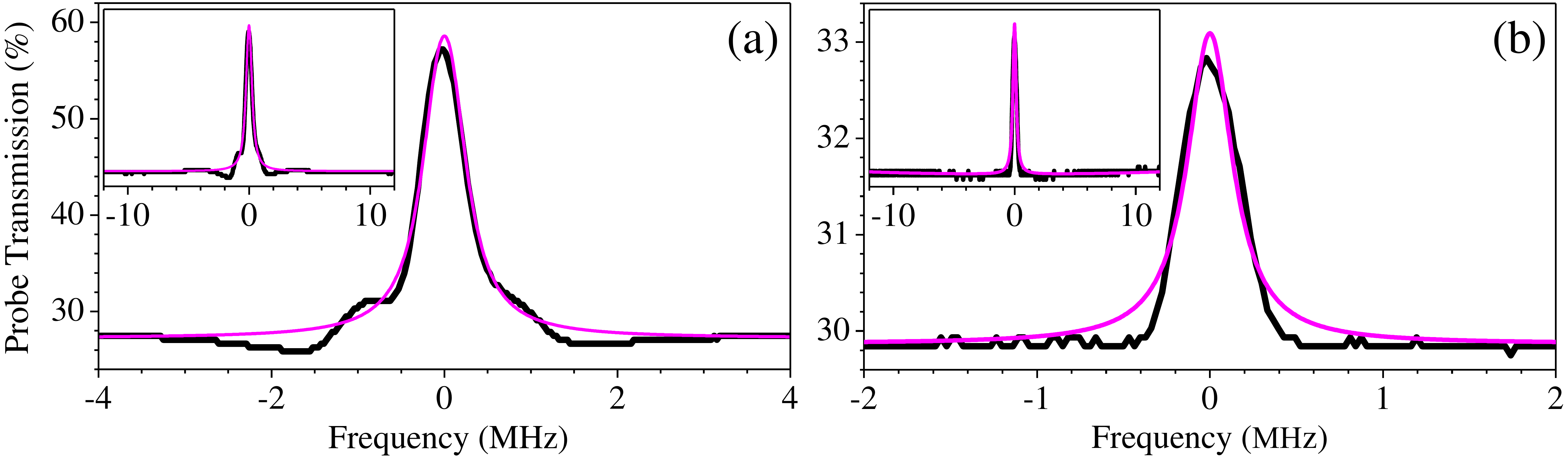}}
	\caption{Representative EIT spectra of the probe field. The powers of the coupling field are 1 mW in (a) and 0.05 mW in (b), that of the HOP field is 18 mW, and that of the input probe field is 50 nW. The pump field had no effect on the spectra. Insets show the same spectra with a larger frequency range. Black lines are the experimental data. Magenta lines represent the best fits of the numerical result calculated with Eq.~(\ref{eq:T_exact}). The best fits determine the FWHMs of 560 kHz in (a) and 300~kHz in (b). The values of OD, coupling Rabi frequency, and decoherence rate are 80, 2.6$\Gamma$, and 0.028$\Gamma$ in (a), and 82, 0.65$\Gamma$, and 0.024$\Gamma$ in (b).
}
	\label{fig:EIT_spectrum}
	\end{figure}
}
\newcommand{\FigFour}{
	\begin{figure}[t]
	\center{\includegraphics[width=0.95\columnwidth]{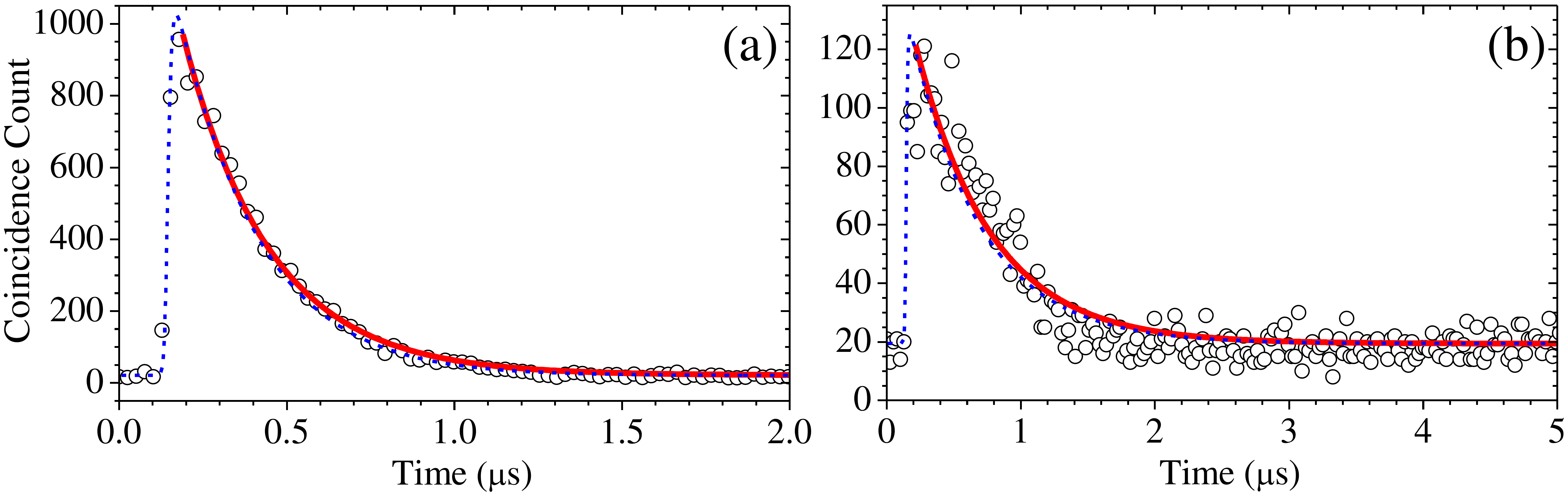}}
	\caption{Representative data of biphoton wave packets measured at the coupling power of 1 mW in (a) and 0.05 mW in (b). The accumulation times are 1200 seconds in (a) and 2400 seconds in (b). The pump and HOP powers are 0.5 mW and 18 mW. Circles are the two-photon coincidence counts, red lines represent the best fits of exponential-decay functions, and blue dashed lines are the theoretical predictions calculated from Eq.~(\ref{eq:biphoton}). The calculation parameters of the theoretical predictions in (a) and (b) are the same as those determined in Figs.~\ref{fig:EIT_spectrum}(a) and \ref{fig:EIT_spectrum}(b), respectively. The $e^{-1}$ time constants of the best fits are 260 ns in (a) and 560 ns in (b), corresponding to the linewidths of 610 kHz and 280 kHz, respectively. We determine the peak signal-to-background ratios of 42 in (a) and 5.4 in (b).}
	\label{fig:biphoton_waverform}
	\end{figure}
}
\newcommand{\FigFive}{
	\begin{figure}[t]
	\center{\includegraphics[width=80mm]{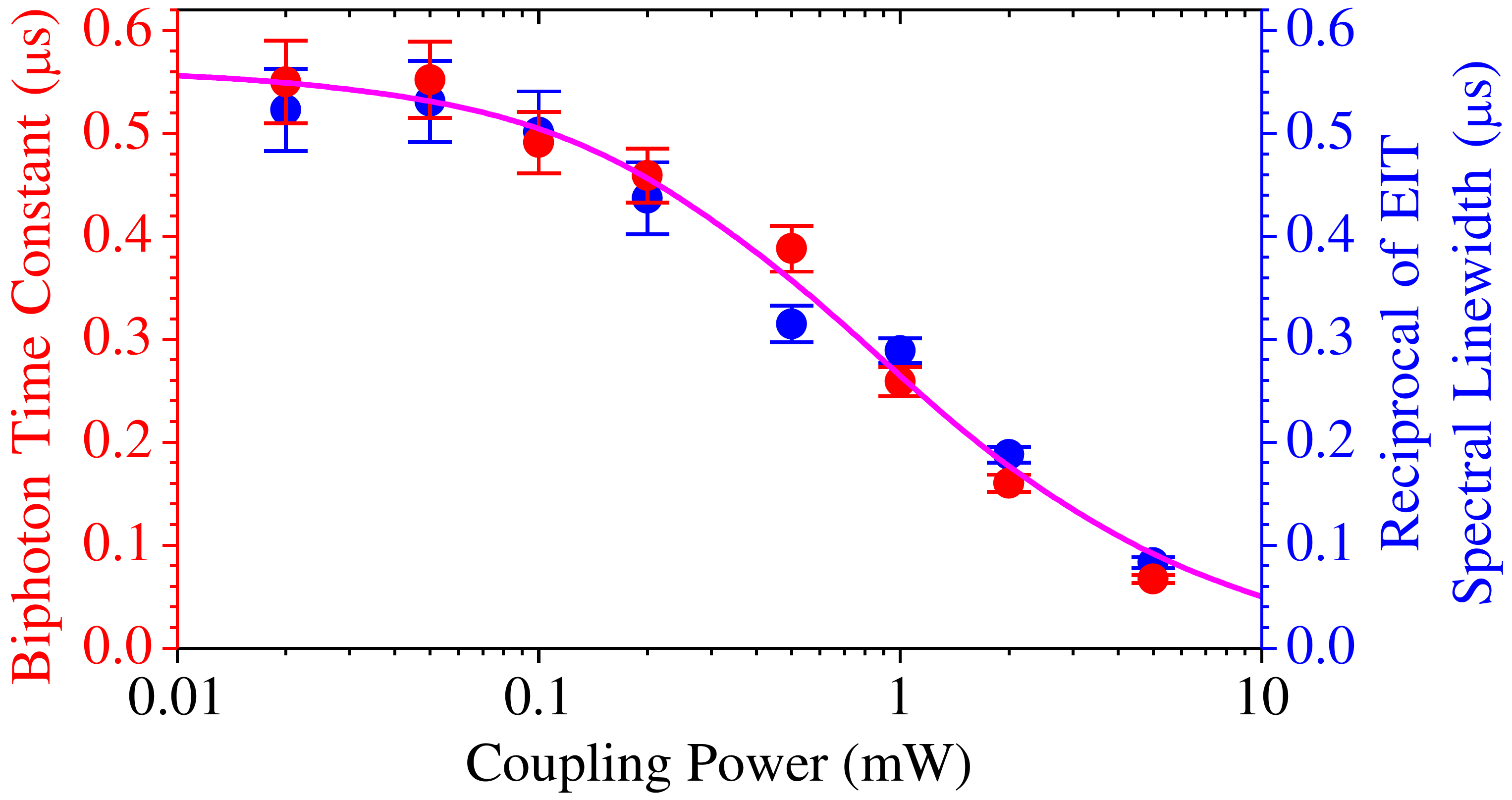}}
	\caption{Time constant of the biphoton wave packet (red) and the reciprocal of FWHM of the EIT spectrum (blue) as functions of the coupling power. Red and blue circles are the experimental data. Magenta line is the theoretical prediction calculated with $\alpha_s$ = 82, $\gamma =$ 0.025$\Gamma$, and $\Omega_c = 2.7\Gamma$$\times$$\sqrt{P_c/(1~{\rm mW})}$ where $P_c$ is the coupling power.}
	\label{fig:temporal_width}
	\end{figure}
}
\newcommand{\FigSix}{
	\begin{figure}[t]
	\center{\includegraphics[width=70mm]{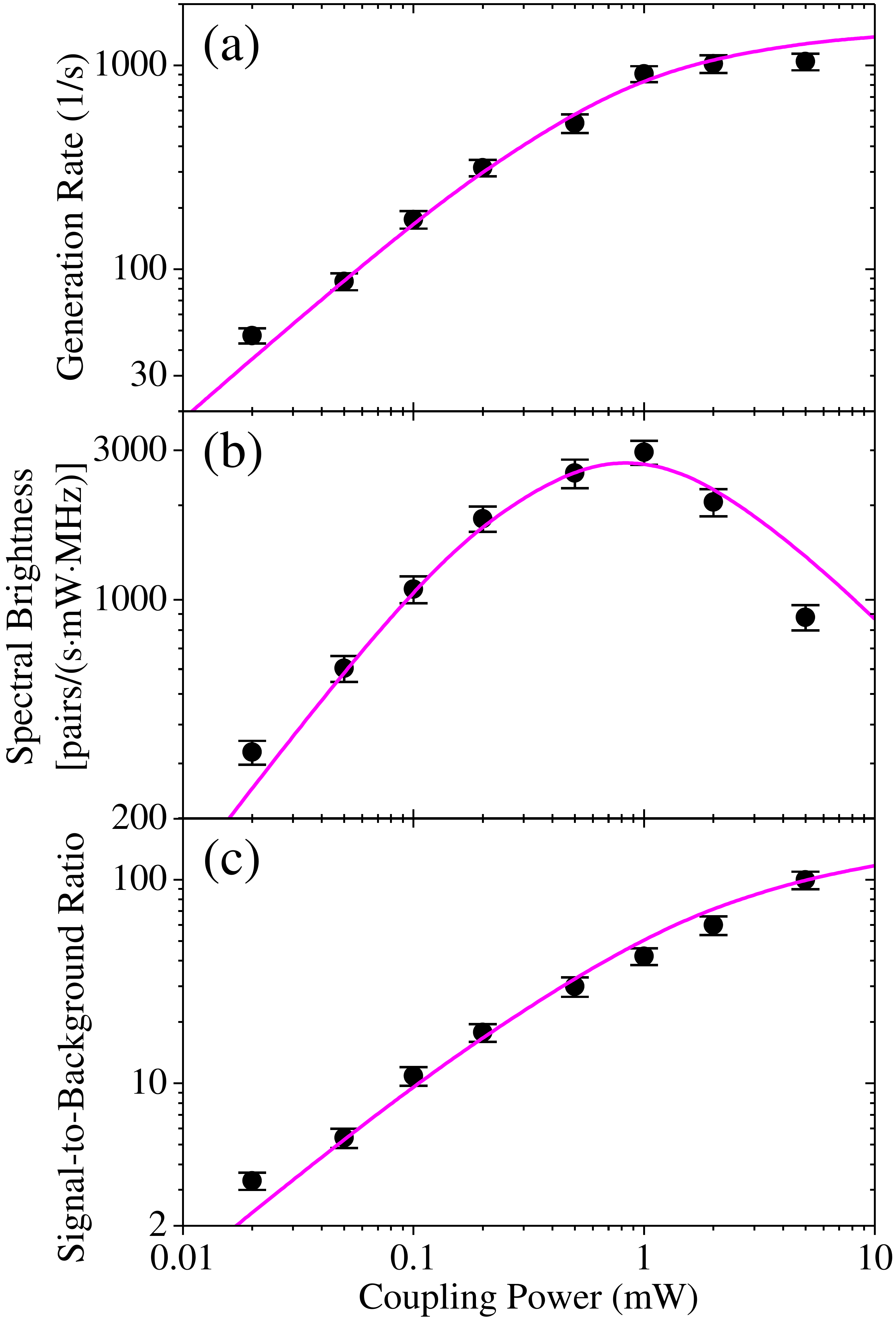}}
	\caption{The generation rate, spectral brightness, and peak signal-to-background ratio of biphotons as functions of the coupling power in (a), (b), and (c), respectively. Black circles are the experimental data. Magenta lines are the theoretical predictions calculated with the same parameters as those in Fig.~\ref{fig:temporal_width}.}
		\label{fig:GR}
	\label{fig:SBR}
	\end{figure}
}
\section{Introduction} \label{sec:introduction}

Photons are superior carriers of information and can keep the carried information intact during the transmission, as they never collide with each other and hardly interact with the environment. Single photons are qubits in long-distance quantum communication \cite{QC1, QC2, QC3, QC4, QC5}. The biphoton is a pair of time-energy entangled single photons \cite{entangled1}. After the first photon of a biphoton pair is detected to start or trigger a quantum operation as the messenger, the second one in the same pair can be conveniently employed in the operation as the heralded qubit. To generate biphotons, the mechanisms of spontaneous parametric down conversion (SPDC) in nonlinear crystals \cite{FirstSPDC, SPDC.Theory, SPDC.Cavity.Review} and spontaneous four-wave mixing (SFWM) in atomic vapors \cite{FirstSFWM, SFWM.Theory, SFWM.dLambda.HotAtom1} are commonly used. 

The photonic qubits of narrower linewidths can make quantum components, such as quantum memories, quantum wavelength converters, and quantum phase gates, have higher efficiencies or success rates \cite{QM1, QM2_OurPRL2013, QM3, QM4, QM5, QM6_OurPRL2018, QM7, QM8, QFC, QPG1, QPG2, QPG3, QPG4}. The SPDC generation can be assisted with an optical cavity to narrow down the linewidth of single photons \CiteSPDC. On the other hand, since the decoherence rate (or coherence time) is typically low (or long) in the cold-atom system, one can also utilize cold atoms in the SFWM process to produce narrow-linewidth biphotons \CiteColdAtom. The SFWM has two types of transition schemes: a double-$\Lambda$ and a ladder schemes. The former is able to produce biphotons with a linewidth of less than 1 MHz \cite{SFWM.dLambda.ColdAtoms7, SFWM.dLambda.ColdAtoms8, SFWM.dLambda.ColdAtoms10}. All the biphotons generated by the latter, in either cold- or hot-atom systems, have linewidths of larger than 10 MHz \CiteLadderType. Thus, we only focus on the SFWM of the double-$\Lambda$ transition scheme in this study.

To date, four groups have reported sources of sub-MHz biphotons. Two groups produced the biphotons with the cavity-assisted SPDC. In Refs.~\cite{SPDC.Cavity10, LongBiphotonSPDC.Cavity}, their biphotons had linewidths of 430 and 265 kHz, and generation rates per linewidth were about 88 and 324 pairs/(s$\cdot$MHz), respectively. Due to the optical cavities, these biphotons are multi-mode and their frequency modes can span a few hundred MHz. The above linewidth and generation rate per linewidth refer to the values of a single frequency mode. Note that some designs of the SPDC's optical cavity can make single-mode biphotons, but currently all of these biphotons have linewidths larger than 1~MHz. The generartion rates or the generation rates per linewidth of these broadband biphoton sources can be as high as the order of $10^6$ pairs/s or $10^5$ pairs/(s$\cdot$MHz) \cite{SPDC.Cavity.Review}. The other two groups generated the biphotons with the SFWM in cold atoms. Their biphotons had linewidths of 250 kHz \cite{SFWM.dLambda.ColdAtoms10} (or 430 kHz \cite{SFWM.dLambda.ColdAtoms7}) and 380 kHz \cite{SFWM.dLambda.ColdAtoms8}, and the generation rates per linewidth were about 470 pairs/(s$\cdot$MHz) \cite{SFWM.dLambda.ColdAtoms7} and 540 pairs/(s$\cdot$MHz) \cite{SFWM.dLambda.ColdAtoms8}.  It is necessary to switch off the mechanism for cooling and trapping the atoms during biphoton generation, and the duty cycle of the generation was 10\%. The above quoted generation rates are averaged over a cycle.

As compared with the biphotons that were generated by the cavity-assisted SPDC in nonlinear crystals or by the SFWM in cold atomic vapors, the biphotons that were previously generated by the SFWM in hot atomic vapors had a broader linewidth and a lower generation rate per linewidth at a given signal-to-background ratio (SBR). The biphotons of hot-atom SFWM in Refs.~\CiteHotAtom\ all had linewidths of larger than 1 MHz. In those studies, the pump and coupling fields counter-propagated, and their propagation directions and the single photons' emission directions had a small separation angle. The counter-propagation scheme was commonly used in the experiments of cold atoms, and did not cause problems since the Doppler effect is negligible and the size of cold atomic clouds is normally small. However, this scheme degrades the linewidth and the spectral brightness (i.e., the generation rate per linewidth per pump power) of biphotons in the experiments of hot atoms. In Refs.~\CiteHotAtom, the best spectral brightness is about 230 pairs/(s$\cdot$mW$\cdot$MHz) \cite{SFWM.dLambda.HotAtom2}.

We generated biphotons with a tunable temporal width from a $^{87}$Rb atomic vapor cell by using the SFWM process. The cell was heated to 38$^{\circ}$C in the experiment. We employed the all-copropagating scheme, instead of the scheme that was commonly used in the previous studies of SFWM with either cold or hot atomic vapors. In the all-copropagating scheme, the pump and coupling fields propagate in the same direction, and completely overlap the emission directions of the Stokes and anti-Stokes photons. The copropagation configuration maintains a good phase-match condition in the SFWM process. The zero angle separation between the strong driving fields and the single photons enables a low decoherence rate in the Doppler-broadened media. Hence, the best spectral brightness after considering the overall collection efficiency in this study reaches 3,000 pairs/(s$\cdot$mW$\cdot$MHz), which is significantly higher than those in the previous studies of the hot-atom SFWM.

In this work, the longest temporal width of the biphoton wave packet is 550$\pm$40 ns, which corresponds to a linewidth of 290$\pm$20 kHz. To our knowledge, this is the narrowest linewidth to date among all single-mode sources of room-temperature and hot media. The maximum two-photon correlation function, $g_{as,s}^{(2)}$, or peak SBR of the 290-kHz biphoton wave packet was 5.4, which violates the Cauchy-Schwartz inequality for classical light by 7.3 folds and clearly demonstrates its nonclassicality. As we tune the temporal width of the biphotons to 260 ns, i.e., the linewidth of 610 kHz, their peak SBR is significantly enhanced to 42, which violates the Cauchy-Schwartz inequality by 440 folds and demonstrates that the heralded single photons have a rather high purity. The 610-kHz biphoton source has the generation rate per linewidth of 1,500 pairs/(s$\cdot$MHz). The biphoton source of hot-atom SFWM not only has the merit of a linewidth tunable for more than an order of magnitude, but also is capable to set to any frequency in a continuous range of 0.6 GHz or larger. Such temporally-long, high-purity, and spectrally-bright biphotons will be very useful in the application of long-distance quantum communication.  

\section{Experimental setup}

Biphotons in the experiment were produced from a paraffin-coated glass cell filled with the vapor of isotopically enriched $^{87}$Rb atoms. The cylindrical cell has the diameter of 25 mm and is 75 mm long. We heated the cell to 38 $^{\circ}$C to maintain the vapor pressure of about 10$^{-6}$ torr or, equivalently, the atomic density of 3.1$\times$10$^{10}$ cm$^{-3}$. Figure~\ref{fig:transition_diagram} shows the relevant energy levels and transitions of SFWM process for the generation of biphotons. Nearly all population was optically pumped to the ground state of $|5S_{1/2},F=2\rangle$. The pump field was red-detuned 2.0 GHz from the transition of $|5S_{1/2},F=2\rangle$ $\rightarrow$ $|5P_{3/2},F''=1\rangle$, and the coupling field drove the transition of $|5S_{1/2},F=1\rangle$ $\rightarrow$ $|5P_{1/2},F'=2\rangle$ resonantly. Due to the vacuum fluctuation, a pair of anti-Stokes and Stokes photons can be spontaneously emitted. The frequencies of the pump field, anti-Stokes photon, coupling field, and Stokes photon form the resonant four-photon transition. The hyperfine optical pumping (HOP) field is not a part of the SFWM process, and was employed to empty the population in $|5S_{1/2},F=1\rangle$.

The coupling field came from a 795-nm external-cavity diode laser (ECDL) of Toptica DL DLC pro 795. We stabilized the coupling frequency with the saturated absorption spectroscopy, and the frequency fluctuation was about $\pm$0.3~MHz. The pump field originated from a homemade 780-nm ECDL. We stabilized the pump frequency with a wavelength meter (HighFinesse WSU-30), and the frequency fluctuation was about $\pm$1.5~MHz. The HOP field came from a laser amplifier (Toptica Boos TA). We seeded the amplifier with the light from a homemade 780-nm bare-diode laser, which was injection-locked by the pump laser light with the offset frequency provided by an electro-optic modulator (EOM). The HOP field was red-detuned 80 MHz from the  transition of $|5S_{1/2},F=1\rangle$ $\rightarrow$ $|5P_{3/2},F''=2\rangle$, and had the same frequency fluctuation as the pump laser.

\FigOne

The experimental setup is shown in Fig.~\ref{fig:experimental_setup}. Both of the pump and coupling fields were linearly polarized in the orthogonal configuration, i.e., they had the $p$ and $s$ polarizations, respectively. We completely overlapped the two fields with a polarization beam splitter. Before the vapor cell, the 780-nm and 795-nm bandpass filters were installed for the pump and coupling fields. These filters together with the 795-nm and 780-nm bandpass filters after the vapor cell prevented the strong pump and coupling light from entering the Stokes and anti-Stokes single-photon counting modules (SPCMs), respectively. The dichroic mirror right after the cell separated the 780-nm and 795-nm light. After the dichroic mirror, the combination of the quarter-wave plate, half-wave plate, and Glan-Thompson polarizer attenuated the pump (or coupling) light by 60~dB (or 48~dB), and allowed a high transmission of the $s$-polarized anti-Stokes photons (or $p$-polarized Stokes photons). A set of two etalons in series further reduced the pump (or coupling) light by about 74~dB (or 88~dB). Considering the above attenuation factors, we estimate that the pump (or coupling) light of 1 mW contributed merely about 100 (or 64) counts/s the anti-Stokes (or Stokes) SPCM. The two SPCMs (Excelitas SPCM-AQRH-13-FC) have the dark-count rates of 140$\pm$5 and 220$\pm$20 counts/s.

Each set of the two etalons has the linewidth of about 35~MHz, and provides a peak transmission of around 37\% (or 42\%) for the anti-Stokes (or Stokes) photons. Because of the high extinction ratios of the etalons, we were able to make the coupling field, Stokes photons, pump field, and anti-Stokes photons propagate in the same direction and completely overlap inside the vapor cell. The anti-Stokes photon propagates with the light speed in vacuum, because of the large one-photon detuning in the Raman process of the pump field and anti-Stokes photon. The Stokes photon is slow light due to the effect of electromagnetically induced transparency (EIT) in the Raman process of the coupling field and Stokes photon. Thus, the anti-Stokes SPCM detected a photon first, triggering the TOF multiscalar, and after some delay time the Stokes SPCM detected another, recorded as a coincidence count. The overall collection efficiencies, including the SPCMs' quantum efficiencies, of the anti-Stokes and Stokes photons were 8.4\% and 13\%, respectively.

\FigTwo

The pump and coupling fields co-propagated inside the vapor cell, and had the $e^{-2}$ full widths of 1.4  and 1.5 mm, respectively. According to the beam width and power, we estimated that the peak intensity of a 0.5-mW pump beam corresponds to the Rabi frequency of 2.0$\Gamma$, where $\Gamma =$ 2$\pi$$\times$6 MHz is the spontaneous decay rate of the excited states. Under the similar estimation, the peak intensity of a 1-mW coupling beam corresponds to the Rabi frequency of 2.9$\Gamma$. The $s$-polarization HOP field has a donut-shaped beam profile with the inner and outer $e^{-2}$ diameters of about 8.5 and 11.5 mm. The donut shape can prevent the HOP field from deteriorating the transmission of Stokes photons, while the HOP field still effectively clears out the population in the state $|2\rangle$ \cite{SFWM.dLambda.HotAtom2}. It propagated in the direction opposite to the pump and coupling's propagation direction. We kept the power of the HOP at 18 mW throughout the experiment. The ratio of the average intensity in the donut area to that in the region of biphoton generation was approximately 30, and the peak Rabi frequency in the donut area is 3.4$\Gamma$.

To study the experimental condition, we measured the EIT spectrum. An input probe field was employed in the measurement. This probe field came from a homemade 795-nm bare-diode laser, which was injection-locked by the coupling laser light with the offset frequency provided by an EOM. We swept the probe frequency across the Stokes transition by tuning the driving frequency of the EOM around 6.835 GHz, i.e., the frequency difference between the two hyperfine levels of the ground state $|5S_{1/2}\rangle$. Because of the injection-lock scheme, fluctuation of the frequency difference between the probe and coupling fields is completely negligible. The input probe field had the $e^{-2}$ full widths of 0.6 mm, and its power was 50 nW. This field is weak enough that it can be treated as the perturbation in the system. We experimentally verified that the probe transmission of the EIT spectrum did not change as the probe power was increased by more than two folds. The input probe field was blocked during the biphoton generation.

\section{Theoretical predictions}

We utilized the time correlation function between the anti-Stokes and Stokes photons, i.e., the biphoton wave packet, to make theoretical predictions, and compared the experimental data with the predictions. The two-photon time correlation is shown below \cite{SFWM.Theory}.
\begin{eqnarray}
\label{eq:biphoton}
	G^{(2)}(\tau)  \!\!\!\!\!\!\!\!\!\! &&=
		\left| \int_{-\infty}^{\infty} d\delta \frac{1}{2\pi} e^{-i\delta\tau}
		\left[ 
			\int_{-\infty}^{\infty} d\omega_D  
			\frac{e^{-\omega_D^2/\Gamma_D^2}}{\sqrt{\pi}\Gamma_D}
			\frac{\sqrt{k_{as} k_s}L}{2} \chi(\delta,\omega_D) 
		\right]
		\right. \\
	&& \!\!\!\!\! \times \left.
		{\rm sinc} \! \left[ 
			\int_{-\infty}^{\infty} d\omega_D  
			\frac{e^{-\omega_D^2/\Gamma_D^2}}{\sqrt{\pi}\Gamma_D}
			\frac{k_s L}{4} \xi(\delta,\omega_D) 
		\right] 
		\exp \left[ 
			i\int_{-\infty}^{\infty} d\omega_D  
			\frac{e^{-\omega_D^2/\Gamma_D^2}}{\sqrt{\pi}\Gamma_D}
			\frac{k_s L}{4} \xi(\delta,\omega_D)
		\right]
		\right|^2,
		\nonumber
\end{eqnarray} 
where $\tau$ is the delay time of the Stokes photon, $\delta$ is the two-photon detuning of the Raman transition between the anti-Stokes photon and pump field (or $-\delta$ is that between the Stokes photon and coupling field), $\omega_D$ is the Doppler shift, $\Gamma_D$ is the Doppler width, $k_{as}$ and $k_s$ are the wave vectors of the two photons, $L$ is the medium length, $\chi(\delta,\omega_D)$ is the cross-susceptibility of the Stokes photon induced by the anti-Stokes photon, and $\xi(\delta,\omega_D)$ is the self-susceptibility of the Stokes photon. The formulas relating the cross-susceptibility and self-susceptibility to the experimental parameters are given by
\begin{eqnarray}
\label{eq:cross_chi} 
	\frac{\sqrt{k_{as} k_s}L}{2} \chi(\delta,\omega_D) \!\!\! &=& \!\!\!
		\frac{\sqrt{\alpha_{as}\alpha_s}\sqrt{\Gamma_3\Gamma_4}}{4} 
		\frac{\Omega_p}{\Delta_p-\omega_D + i\Gamma_4/2} 
		\frac{\Omega_c}{\Omega_c^2-4(\delta+i\gamma)(\delta+\omega_D+i\Gamma_3/2)},~~ 
		\\
\label{eq:self_chi}
	\frac{k_s L}{4} \xi(\delta,\omega_D) \!\!\! &=& \!\!\! \frac{\alpha_s\Gamma_3}{2} 
		\frac{\delta+i\gamma}{\Omega_c^2-4(\delta+i\gamma)(\delta+\omega_D+i\Gamma_3/2)},
\end{eqnarray}
where $\alpha_s = n \sigma_s L$ ($n$ is the atomic density and $\sigma_s$ is the resonant absorption cross section of the Stokes transition) represents the optical depth of the entire atoms interacting with the Stokes photon resonantly, $\alpha_{as}$ means the similar optical depth of the anti-Stokes transition, $\Omega_p$ and $\Omega_c$ are the Rabi frequencies of the pump and coupling fields, $\Gamma_3$ and $\Gamma_4$ are the spontaneous decay rates of the excited states (i.e., $|3\rangle$ and $|4\rangle$ in Fig.~\ref{fig:transition_diagram}) in the Stokes and anti-Stokes transitions, respectively, $\Delta_p$ is the detuning of the pump field, and $\gamma$ is the dephasing rate of the ground-state coherence, i.e., the decoherence rate. Since the difference between $\Gamma_3$ and $\Gamma_4$ is merely about 5\% in our case, we neglect the difference and set $\Gamma_3 = \Gamma_4 \equiv \Gamma = 2\pi\times6~{\rm MHz}$ in this work. Due to the temperature of the vapor cell being 38$^{\circ}$C, $\Gamma_D =$ 54$\Gamma$ in the calculation. Equation~(\ref{eq:biphoton}) is the consequence of the four-photon resonance, i.e., $\omega_c+\omega_p=\omega_{as}+\omega_{s}$, and it reveals the time-energy entanglement \cite{entangled1}.

We measured the EIT spectrum to characterize the experimental parameters  \cite{OurSR2018, OurPRA2019}. The theoretical EIT spectrum is given by the self-susceptibility $k_s L\xi(\delta,\omega_D)$ of the Stokes photon or, equivalently, the classical probe field. Considering a Doppler-broadened medium, the imaginary part of $k_s L\xi(\delta,\omega_D)$ is integrated over all the velocity groups. Thus, we obtain the transmission, $T$, of the probe field as the following:
\begin{equation}
	T = {\rm exp}\left\{
		-\int_{-\infty}^{\infty} d\omega_D 
		\frac{ e^{-\omega_D^2/\Gamma_D^2}}{\sqrt{\pi}\Gamma_D}
		~{\rm Im}[k_s L\xi(\delta,\omega_D)] \right\}.
\label{eq:T_exact}
\end{equation}
In the experiment, we fitted the measured EIT spectra with the calculation results of the above formula. The best fits determine the experimental parameters of $\alpha_s$, $\Omega_c$, and $\gamma$.

\section{Results and discussion} \label{sec:results}

Once the biphoton is generated, the anti-Stokes photon with the light speed in vacuum quickly exits the medium, and the Stokes photon is the slow light and propagates in the medium under the presence of the coupling and pump fields. Since the pump field was far detuned, it had a negligible effect on the Stokes photon. The temporal profile of the probability of detecting the Stokes photon, i.e., the biphoton wave packet, is mainly determined by the EIT effect. Consequently, the measured EIT spectrum can reveal the experimental condition of optical depth ($\alpha_s$), coupling Rabi frequency ($\Omega_c$), and decoherence rate ($\gamma$) for the theoretical calculation to predict the biphoton wave packet. We measured the spectra with an input probe field under the presence of the coupling, HOP, and pump fields. The input probe field is weak enough that it can be treated as the perturbation in the system. The coupling power was varied and set to either 0.02, 0.05, 0.1, 0.2, 0.5, 1, 2, or 5 mW. Since we intended to make the experimental condition as close to the biphoton generation as possible, the HOP and pump fields were also present in the spectrum measurement. Nevertheless, at each coupling power the spectra with and without the pump field had no difference. The black lines in Figs.~\ref{fig:EIT_spectrum}(a) and \ref{fig:EIT_spectrum}(b) are the representative EIT spectra at the coupling powers of 1 and 0.05 mW.

\FigThree

We fitted the EIT spectra of different coupling powers with the numerical result calculated from Eq.~(\ref{eq:T_exact}). The value of $\alpha_s$ is uniquely determined by the baseline transmission of the spectrum. At a given $\alpha_s$, the peak transmission and width of the EIT window resolve the values of $\Omega_c$ and $\gamma$. Thus, the values of $\alpha_s$, $\Omega_c$, and $\gamma$ are unambiguously determined by the best fits of the EIT spectra. The magenta lines in Figs.~\ref{fig:EIT_spectrum}(a) and \ref{fig:EIT_spectrum}(b) are the best fits. In Fig.~\ref{fig:EIT_spectrum}(a), the experimental spectrum shows small peaks and dips on the two sides of the central EIT peak, which are not present in the best fit calculated from Eq.~(\ref{eq:T_exact}). These minor peaks and dips were caused by the residual light of the HOP field in its hollow region. As we turned off the HOP field, they disappeared. In Fig.~\ref{fig:EIT_spectrum}(b), the part of the experimental spectrum near the baseline deviates from the best fit. This deviation was also caused by the HOP field.

We systematically measured the biphoton data at different coupling powers of 0.02, 0.05, 0.1, 0.2, 0.5, 1, 2, and 5 mW. The temperature of the atomic vapor cell, i.e., the optical depth (OD), was varied to maximize the SBR at the coupling power of 2 mW. The optimum temperature is 38$^{\circ}$C. Figures~\ref{fig:biphoton_waverform}(a) and \ref{fig:biphoton_waverform}(b) show the representative biphoton wave packets, i.e., the coincidence count as a function of the delay time between the Stokes and anti-Stokes photons, taken at the coupling powers of 1 and 0.05 mW. The success probabilities of detecting a Stokes photon upon an anti-Stokes-photon trigger are $0.88\%$ and $0.093\%$ in Figs. \ref{fig:biphoton_waverform}(a) and \ref{fig:biphoton_waverform}(b), respectively. The bin time of data points is 25.6 ns. At the coupling power of 0.05 mW, the pump power was optimized for the best signal-to-background ratio (SBR), and the HOP power was optimized for the longest temporal width without degrading the SBR. We used the same pump and HOP powers for all the biphoton measurements. Since the pump power was kept the same in the biphoton generation, the detection rates (excluding the SPCM's dark counts) of the anti-Stokes photons at different coupling powers were all around 840$\pm$50 counts/s. The fluctuation of $\pm$6\% was mainly comes from the variation of the collection efficiency of anti-Stokes photons caused by the drift of the etalon's temperature.

The biphoton wave packet behaves nearly like an exponential-decay function, i.e.,  $y(x) = y_0 + S\exp[-(x-x_0)/\tau]$, as shown by the red lines in Fig.~\ref{fig:biphoton_waverform}. We resolved $y_0$ (the baseline) from the data first, and set $x_0$ to 200~ns and fitted the data with the fitting parameters of $S$ (the amplitude) and $\tau$ (the time constant). In each biphoton wave packet, we determine the linewidth from $\tau^{-1}/(2\pi)$ of the best fit and calculate the SBR from the ratio of $S$ to $y_0$. We further use Eq.~(\ref{eq:biphoton}) to calculate the theoretical predictions of biphoton data as shown by the blue dashed lines in Figs.~\ref{fig:biphoton_waverform}(a) and \ref{fig:biphoton_waverform}(b). Two Lorentzian functions of the FWHMs of 45 and 60 MHz, corresponding to the etalons' spectral profiles, were added to Eq.~(\ref{eq:biphoton}). The calculation parameters of OD, coupling Rabi frequency, and decoherence rate in Figs.~\ref{fig:biphoton_waverform}(a) and \ref{fig:biphoton_waverform}(b) are the same as those determined in Figs.~\ref{fig:EIT_spectrum}(a) and \ref{fig:EIT_spectrum}(b), respectively. We optimized the amplitudes and baselines of the predictions to match the data. As shown by Fig.~\ref{fig:biphoton_waverform}, the consistency between the theoretical predictions and experimental data (as well as their best fits) is satisfactory.

\FigFour

To study the relation between the temporal width of the biphoton wave packet and the linewidth of the corresponding EIT spectrum, we plot the biphoton time constant ($\tau$) and the reciprocal of the EIT FWHM ($\Gamma_{\rm EIT}^{-1}$) as functions of the coupling power in Fig.~\ref{fig:temporal_width}. Although there exist some discrepancies between the two data sets, their overall behaviors are very similar. A smaller coupling power results in a longer temporal width of the biphoton wave packet or a narrower linewidth of the EIT spectrum. Furthermore, the biphoton's temporal width (or EIT linewidth) asymptotically approaches an upper (or lower) limit, which is theoretically equal to $(2\gamma)^{-1}$. The temporal width (or linewidth) of the biphoton wave packet is limited to 550$\pm$40 ns (or 290$\pm$20 kHz) measured at the coupling power of 0.05 mW or less. Using the value of $\gamma$ determined from the corresponding EIT spectrum, $(2\gamma)^{-1} \approx$ 530 ns (or  $\gamma/\pi \approx$ 300 kHz). The linewidth of EIT spectrum measured with classical light can be a good indicator of the temporal width or spectral linewidth of biphoton wave packet.

We further employed Eq.~(\ref{eq:biphoton}) to predict the biphoton wave packet of each coupling power, and compared the experimental data with the theoretical predictions. In the numerical calculation of Eq.~(\ref{eq:biphoton}), the parameters of the OD ($\alpha_s$), coupling Rabi frequency ($\Omega_c$), and decoherence rate ($\gamma$) were set under the following consideration. As mentioned earlier, the best fit of the EIT spectrum measured at each coupling power had determined a set of $\alpha_s$, $\Omega_c$, and $\gamma$. First of all, we fitted the data points of $\Omega_c^2$ versus the coupling power, $P_c$, with a straight line of zero interception. The best fit gives the relation of $\Omega_c = 2.7\Gamma \sqrt{P_c/(1~{\rm mW})}$. Secondly, the best-fit value of $\gamma$ slightly increased with the coupling power (see the examples in Fig.~\ref{fig:EIT_spectrum}). The biphoton's temporal width of a small coupling power is sensitive to $\gamma$, but that of a large coupling power is not. Hence, we chose the average value of $\gamma$'s of the three smallest coupling powers in the calculation. Finally, the values of $\alpha_s$ determined from the experimental spectra vary from 79 to 84, and the variation affects the prediction little. We set $\alpha_s$ to the intermediate value, i.e., 82. The other parameters of the pump Rabi frequency ($\Omega_p$), pump detuning ($\Delta_p$), and anti-Stokes photon's OD ($\alpha_{as}$) do not affect the shape and width of the biphoton wave packet, and only change its overall magnitude. In Fig.~\ref{fig:temporal_width}, the magenta line is the theoretical prediction calculated with $\alpha_s = 82$, $\gamma = 0.025$$\Gamma$, and $\Omega_c = 2.7\Gamma \sqrt{P_c/(1~{\rm mW})}$. The consistency between the experimental data and theoretical predictions is satisfactory. 

\FigFive 

The generation rate is an important figure of merit of a biphoton source. We show how the coupling power or equivalently $\Omega_c^2$ affects the biphoton generation rate in Fig.~\ref{fig:GR}(a). The black circles are the experimental data which are the results of the coincidence count per second or the detection rate divided by the product of the collection efficiencies of the anti-Stokes and Stokes photons. As $\Omega_c^2$ increases, the generation rate is enhanced accordingly, but eventually gets saturation. The observation is in agreement with the situation that the Stokes photon propagates in the EIT medium like slow light and a larger $\Omega_c^2$ makes the transmission higher. At a very large $\Omega_c^2$, all the generated Stokes photons can move out of the medium with a little attenuation and the generation rate saturates. To simulate the generation rate as a function of the coupling power, the same predictions of the biphoton wave packet in Fig.~\ref{fig:temporal_width} were used. The area below the wave packet is proportional to the generation rate. We multiplied the area by a normalization factor to obtain the predictions of generation rate as shown by the magenta line in Fig.~\ref{fig:GR}(a). The normalization factor minimizes the standard deviation between the experimental data and the predictions. Other than getting saturation a little faster, the experimental data behave similarly to the theoretical predictions. 

The spectral brightness is defined as the generation rate per pump power per spectral linewidth. We plot the spectral brightness as a function of the coupling power in Fig.~\ref{fig:GR}(b).  The circles are the experimental data. The magenta line represents the theoretical predictions which are the predictions in Fig.~\ref{fig:GR}(a) divided by the product of the prediction in Fig.~\ref{fig:temporal_width} and the pump power of 0.5 mW. At the coupling power of 1~mW, the biphoton source reaches the maximum spectral brightness of 3,000 pairs/(s$\cdot$mW$\cdot$MHz). As the coupling power gets larger than 1 mW, the generation rate becomes saturated and increases a little, but the linewidth still increases prominently. The maximum of the spectral brightness locating around 1 mW is expected. 

\FigSix

The SBR in the biphoton wave packet can provide the information of the two-photon correlation function $g_{as,s}^{(2)}$ between the anti-Stokes and Stokes photons. Since the biphoton wave packet is an exponential-decay function, the SBR ($= S/y_0$ mention earlier) is equal to the maximum $g_{as,s}^{(2)}$. Figure~\ref{fig:SBR}(c) shows the SBR as a function of the coupling power. The black circles are the experimental data, and the magenta lines represent the theoretical predictions. To make the predictions of SBR, we utilized Eq.~(\ref{eq:biphoton}) to calculate the biphoton wave packets with the same parameters as those in Fig.~\ref{fig:temporal_width}. The effect of the rise time of 35 ns observed in the experimental data is also included in the calculation. Next, we got the ratio of the peak of the theoretical waveform to the background count rate. This background count rate, $B$, was measured regardless of the anti-Stokes photons, and relates to the coupling power, $P_c$, as $B = 240 + 320 [P_c/(1~{\rm mW})]^{0.53}$ counts/s. Finally, we multiplied the above ratio by a normalization factor to obtain the predicted SBR. In Fig.~\ref{fig:SBR}(c), the behavior of the experimental data agrees with that of the theoretical predictions.

In this work, the maximum $g_{as,s}^{(2)}$ of the biphotons of the narrowest linewidth, measured at the coupling power of 0.05 mW, is 5.4. The result violates the Cauchy-Schwartz inequality for classical light, i.e., $[g_{as,s}^{(2)}]^2 / [g_{as,as}^{(2)} \cdot g_{s,s}^{(2)}] \leq 1$, by 7 folds, and clearly demonstrates nonclassicality of these 290-kHz biphotons. To test the inequality, we set the values of $g_{as,as}^{(2)}$ and $g_{s,s}^{(2)}$ to 2 \cite{auto_2}, which is close to the measured values in Refs.~\cite{SFWM.dLambda.HotAtom1, SFWM.dLambda.HotAtom6}. As we set the coupling power to 2 mW, the biphotons had the temporal width of 160~ns, i.e., the linewidth of just below 1 MHz. The SBR of these 1-MHz biphotons was significantly enhanced to 60 which violates the Cauchy-Schwartz inequality by 900 folds, exhibiting high purity. At the coupling power of 1 mW and the pump power of 0.5 mW, the 610-kHz biphoton source had the SBR of 42 and the generation rate per linewidth of 1,500 pairs/(s$\cdot$MHz). By changing the pump power from 0.5 to 8 mW, we were able to enhance the generation rate per linewidth of the biphoton source to 2.3$\times$10$^4$ pairs/(s$\cdot$MHz) at expense of the SBR being reduced to 6.7 and the linewidth being slightly increased to 670 kHz.

\section{Prospects and conclusion}

We have generated biphotons from the $^{87}$Rb atomic vapor cell by using the SFWM process, and systematically studied the temporal width or spectral linewidth, the generation rate, the spectral brightness, and the maximum two-photon correlation function $g_{as,s}^{(2)}$ or SBR as functions of the coupling power or equivalently $\Omega_c^2$. The coupling power was varied from 0.02 to 5 mW. The consistency between the experimental data and theoretical predictions that were calculated from Eq.~(\ref{eq:biphoton}) is satisfactory.

Here, we employed the all-copropagating scheme. The scheme not only maintains a good phase-match condition in the SFWM, but also enables a low decoherence rate in the Doppler-broadened medium. Consequently, we have been able to generate sub-MHz biphotons, which could previously not be achieved with hot atom vapors. In addition, the spectral brightness of our sub-MHz biphoton source is significantly higher than those of the early hot-atom SFWM studies. Please note that a longer vapor cell can have a larger OD, resulting in a higher biphoton generation rate. However, the OD-enhancement effect is cancelled out by a larger phase mismatch in the counter-propagation scheme due to a longer optical path. The all-copropagating scheme demonstrated here can maintain a good phase-match condition regardless of the length of the vapor cell. Thus, one can use a longer cell to achieve a higher generation rate or spectral brightness. The results of the generation rate and spectral brightness presented here can be scaled up with the square of the OD or cell length.

The linewidth-tunable biphotons produced here can compete with those generated by the cavity-assisted SPDC and by the cold-atom SFWM. We were able to tune the linewidth down to 290$\pm$20 kHz. Previously, the single-mode biphoton sources of cavity-assisted SPDC all had linewidths of larger than 1 MHz \cite{SPDC.Cavity.Review}, and those of cold-atom SFWM had the narrowest linewidth of 250 kHz \cite{SFWM.dLambda.ColdAtoms10}. The multi-mode biphoton sources can have a frequency mode span of a few hundred MHz, and the linewidth of a mode can be as narrow as 265 kHz \cite{LongBiphotonSPDC.Cavity}. Our biphoton source of 610-kHz linewidth had a generation rate per linewidth of 1,500 pairs/(s$\cdot$MHz) at an SBR of 42. As we increased the pump power by 16 folds, the SBR became 6.7 and the generation rate per linewidth was enhanced to 2.3$\times$10$^4$ pairs/(s$\cdot$MHz).

In conclusion, biphotons are pairs of time-energy entangled single photons and can be employed as heralded photonic qubits in long-distance quantum communication. The biphoton source of hot-atom SFWM not only has the merit of a linewidth tunable for more than an order of magnitude, but also is capable to set to any frequency in a continuous range of 0.6 GHz or larger. Since the sub-MHz, high-purity, and spectrally-bright biphotons produced in this work will be the versatile and powerful carriers of information, the results of this work have become an important milestone in the quantum technology utilizing photonic qubits.

\section*{Acknowledgments}
This work was supported by Grant Nos.~108-2639-M-007-001-ASP and 109-2639-M-007-001-ASP of the Ministry of Science and Technology, Taiwan.

\section*{Disclosures}
The authors declare no conflicts of interest.


\end{document}